\documentclass[twocolumn,amsmath,showpacs,aps,pre]{revtex4}
\usepackage[dvips]{graphicx}
\usepackage{dcolumn}
\begin{document}
%
%

\title{Multifractal spectra of mean first-passage time 
       distributions in disordered chains}

\author{Pedro A. Pury}
\email{pury@famaf.unc.edu.ar}
\affiliation{
Facultad de Matem\'atica, Astronom\'\i a y F\'\i sica, \\
Universidad Nacional de C\'ordoba, \\
Ciudad Universitaria, 5000 C\'ordoba, Argentina}

\author{Manuel O. C\'aceres}
\email{caceres@cab.cnea.gov.ar}
\affiliation{
Centro At\'omico Bariloche, Instituto Balseiro, and\\
Consejo Nacional de Investigaciones Cient\'\i ficas y T\'ecnicas, \\
8400 San Carlos de Bariloche, R\'\i o Negro, Argentina}

\date{Physical Review E - Received 24 February 2003}

%
%

\begin{abstract}
The multifractal characterization of the distribution over disorder
of the mean first-passage time in a finite chain is revisited.
Both, absorbing-absorbing and reflecting-absorbing boundaries are
considered.
Two models of dichotomic disorder are compared and our analysis
clarifies the origin of the multifractality.
The phenomenon is only present when the diffusion is anomalous.
\end{abstract}

\maketitle

\section{Introduction}
\label{sec:intro}

In the past two decades, a great effort has been devoted to the study
of diffusion and transport in disordered media by models based on
random walks~\cite{ABSO81,Reviews,HB91}.
Basically, the dynamics of the system can be described by a 
master equation for a particular probability distribution. 
Two alternative ways are usually employed. The first one is based 
on the probability $P(r,t)$ that the walker is on site $r$ at time 
$t$ when starting from the origin at $t=0$. The second way consists
of analyzing the statistics of the exit time from a given region.
In both cases, there are exact enumeration techniques which enable
us to calculate the corresponding observable in all the possible 
disorder configurations. Thus, we can numerically reckon the 
distribution over disorder of the probability $P(r,t)$, 
or the moments $\left< P^q(r,t) \right>$ of this distribution
\cite{BHHR87,RBH88,BHR90,HB91,Rom92,EBHR93,EHW94}.
In an analogous manner, we can numerically evaluate the 
distribution of the mean first-passage time (MFPT) over disorder,
or its moments~\cite{MD90,van91,MRK94,WGM95,MKG96,MGK96,Kim},
%
\begin{equation}
M(q) = \int_0^{\infty} T^q \,\rho(T) \,dT \,,
\label{M}
\end{equation}
where $\rho(T)$ is the distribution over disorder and 
$q$ is a real number not necessarily integer. 
The crucial point of diffusion in disordered media is that 
transport can be {\em anomalous}, i.e., the mean square 
displacement of a random walk scales with time as 
$\left< r^2 \right> = D \,t^{2/d_w}$,
where $d_w > 2$ is the anomalous diffusion exponent.
A relevant property of the anomalous diffusion is that 
it leads to broad distributions. Their moments cannot 
be described by a single exponent but an infinite hierarchy 
of exponents is needed to characterize them~\cite{EHW94,WGM95}.
This fact enables us to study the distributions over disorder 
with the multifractal formalism~\cite{HJK+86,Fed88}.

Random walks on random fractals (such as the infinite percolation
cluster at criticality) are processes with anomalous diffusion,
which are characterized by logarithmically broad distribution 
functions reflecting an underlying multifractal structure
\cite{BHHR87,BHR90,MD90,HB91,Rom92,EBHR93}.
An important issue in these systems is the nature of the 
multifractality based on the dynamical process or in the
steady state (as the voltage drop distribution in percolation).
For the voltage drop distribution, the multifractal behavior is well
established for all values of $q$~\cite{HB91}; whereas for the
dynamical diffusive process the multifractality only appears in 
a range of values of $q$~\cite{EBHR93}.
Another property of these processes is that while the multifractal
behavior appears in the distribution of the probability $P(r,t)$, 
it is not present in the mean number of distinct sites visited 
by a particle diffusing on the percolating cluster~\cite{Gallos}.
Another dynamical process with ``ultra-anomalous'' slow motion
is the Sinai model~\cite{sinai} for diffusion in a linear chain 
in the presence of random fields. 
For this model, the mean square displacement of a random walk
follows asymptotically $\left< r^2 \right> \approx \ln^4t$. 
Several studies~\cite{RBH88,van91,MRK94,MKG96,MGK96,HM00} 
have established the multifractal properties of the model.
However, the origin of the multifractality in this process
is yet an open question. 

In this work, we revisit the Sinai model and we address the last 
question. Particularly, we compare the behavior of the MFPT
distributions over disorder and its moments for two processes 
with dichotomic disorder. One of them is the Sinai model and 
the other is a nonanomalous biased random walk in a finite 
disordered chain.
The outline of the paper is as follows. In Sec.~\ref{sec:mfpt},
we present expressions for the MFPT for a given realization of the
(quenched) disorder in the chain. The description of the models of
disorder, employed in our studies, is given in Sec.~\ref{sec:models}.
The moments of the MFPT distribution over disorder diverge with the
system's size. These divergences are characterized by the scaling
exponents $\xi(q)$ in Sec.~\ref{sec:multi}, where we employ the
multifractal formalism~\cite{HJK+86,Fed88} to calculate the 
exponents $\tau(q)$ of the corresponding partition function. 
Also, the generalized Renyi dimensions $D(q)$, and the spectra
$f(\alpha)$ are given in that section. 
Finally, in Sec.~\ref{sec:fin}, we briefly summarize the results 
of our work.

\section{Mean first-passage time in quenched disordered chains}
\label{sec:mfpt}

We consider the continuous time dynamics of a random walk on a
discrete one-dimensional lattice with nearest neighbor hopping.
The walker jumps from site $n$ to site $n+1$ with transition
probability per unit time $w^+_n$, or to site $n-1$ with transition 
probability per unit time $w^-_n$. We are concerned with the exit time
of the walker from the finite interval $D = [-M,L]$ on the chain,
with at least one absorbing end. The average of the survival time
until the absorption, over realizations of the random walk, 
is the MFPT.
Recently, we have obtained a general exact expression for the MFPT
for a fixed set of transition probabilities $\{ w^\pm_j \}$
\cite{Pury03}. Let $T_n$ denote the MFPT if the walker initially 
began at site $n \in D$. For an interval with both absorbing 
extremes $(w^+_{-(M+1)} = w^-_{L+1} = 0)$ we get
%
\begin{eqnarray}
T_n &=&
\frac
{\displaystyle
{1+\sum_{k=-M}^{n-1}\prod_{j=-M}^{k}\;\frac{w^-_j}{w^+_j}}}
{\displaystyle
{1 +\sum_{k=-M}^{L} \prod_{j=-M}^{k}\;\frac{w^-_j}{w^+_j}}}
\nonumber \\
&& \times
\left(
\displaystyle\sum_{k=-M}^{L} \,\frac{1}{w^+_k} +
\displaystyle\sum_{k=-M}^{L-1} \,\frac{1}{w^+_k} \,
\sum_{i=k+1}^{L} \prod_{j=k+1}^{i} \;\frac{w^-_j}{w^+_j}
\right)
\nonumber \\
&& -
\left(
\displaystyle\sum_{k=-M}^{n-1} \,\frac{1}{w^+_k} +
\displaystyle\sum_{k=-M}^{n-2} \,\frac{1}{w^+_k} \,
\sum_{i=k+1}^{n-1} \prod_{j=k+1}^{i} \;\frac{w^-_j}{w^+_j}
\right)
\;.
\nonumber \\
\label{MFPT:aa}
\end{eqnarray}
%

In this work we fix the starting point $n=0$ (defining $T \equiv T_0$)
and we consider two possible boundary conditions. 
The first one is the interval $[-L,L]$ with absorbing-absorbing (AA)
extremes. The total number of sites, $N_s$, in the interval is
$2L+1$. The MFPT for this case is given by Eq.~(\ref{MFPT:aa})
taking $M = L$. The second case is the interval $[0,L]$ with
reflecting-absorbing (RA) ends. Here, $N_s = L+1$ and  
from Eq.~(\ref{MFPT:aa}), taking $n = 0$, $M = 0$, and $w^-_0 = 0$,
we immediately obtain for the MFPT~\cite{Pury03},
\begin{equation}
T =
\displaystyle\sum_{k=0}^{L} \,\frac{1}{w^+_k} +
\displaystyle\sum_{k=0}^{L-1} \,\frac{1}{w^+_k} \,
\sum_{i=k+1}^{L} \prod_{j=k+1}^{i} \;\frac{w^-_j}{w^+_j}
\;.
\label{MFPT:ra_0}
\end{equation}
The effect of the reflecting boundary is the striking simplification
of the structure of the equation for the MFPT. This fact leads us 
to consider also the AA boundary conditions in the problem of 
multifractality of the MFPT distribution over disorder.

In finite discrete systems, we can enumerate all the configurations
${\cal N}$ of disorder. We denote by $T_{(i)}$ the MFPT for the 
$i$th realization of the quenched disorder. We stress that all the
configurations of disorder are equally probable. However, the
resulting values $T_{(i)}$ are distributed by $\rho(T)$,
the MFPT distribution over disorder.
Thus, we can compute exactly the moments of the MFPT distribution,
given the set of values $\{ T_{(i)}, \,i = 1, \dots, {\cal N} \}$, 
from the definition given by Eq.~(\ref{M}),
\begin{equation}
M(q) \equiv \frac{1}{\cal N} \;\sum_{i=1}^{\cal N} \; T_{(i)}^q \,.
\label{M(q)}
\end{equation}
In particular, the MFPT averaged over disorder results in $M(1)$.
For dichotomic models of disorder, in a chain with $N_s$ sites, 
there are ${\cal N} = 2^{N_s}$ possible realizations of the random
lattice, which can be easily enumerated. The set of values
$\{T_{(i)}, \,i= 1,\dots, {\cal N} \}$ can be exactly
calculated employing  expression~(\ref{MFPT:aa}) for the AA boundary
conditions, or by Eq.~(\ref{MFPT:ra_0}) for the RA extremes.

\section{Models of disordered chains}
\label{sec:models}

Now, we assume that the hopping probabilities $w^\pm_j$ are strictly
positive random variables, chosen independently from site to site and
identically distributed. Additionaly, we admit that the site
transition probabilities are not necessarily symmetric in the sense
that $w^+_j \neq w^-_j$. Thus, we can incorporate the effects of bias 
in to the chain by external fields.

We select the first distribution for the transition probabilities
in such a way that they satisfy the Sinai condition~\cite{sinai},
namely, the random variable $\ln \left( w^+_j/w^-_j \right)$ has zero
mean and finite variance $\sigma^2$. Thus, we consider a dichotomic
model by defining $w^-_j = 1 - w^+_j$ and prescribing $w^+_j$ to be
equal to $1/2 \pm \epsilon$ with equal probabilities. The parameter
$\epsilon$ measures the strength of disorder, and can take values
between $0$ and $1/2$. $\epsilon = 0$ corresponds to a simple
homogeneous random walk, and for $\epsilon > 0$ we get a disordered
random walk with local bias (random field). It is easily verified
that the above prescriptions satisfy the Sinai condition.
Particularly, we obtain
$\sigma^2 = \ln^2 \gamma(\epsilon)$, where 
$\gamma(\epsilon) \equiv (1 + 2 \,\epsilon) / (1 - 2 \,\epsilon)$.
In the asymptotic limit $\epsilon \rightarrow 1/2$, the variance
diverges. 
The MFPT averaged over disorder for the dichotomic Sinai model 
diverges as $\beta(\epsilon)^{N_s}$ for $N_s \rightarrow \infty$,
where 
$\beta(\epsilon) \equiv (1 + 4\,\epsilon)/(1 - 4\,\epsilon)$
\cite{MK89}, whereas the typical value of the first-passage time,
defined as $\exp (\left< \ln T \right>)$, diverges slower, as 
$\exp\left( \sigma \sqrt{\pi N_s / 2} \right)$~\cite{NG88}.
Here, the brackets $\left< \cdots \right>$ denote the average 
over disorder.
Therefore, the distribution of the MFPT over the Sinai disorder 
has a power-law tail~\cite{tail}.

\begin{figure}
\includegraphics[clip,width=0.45\textwidth]{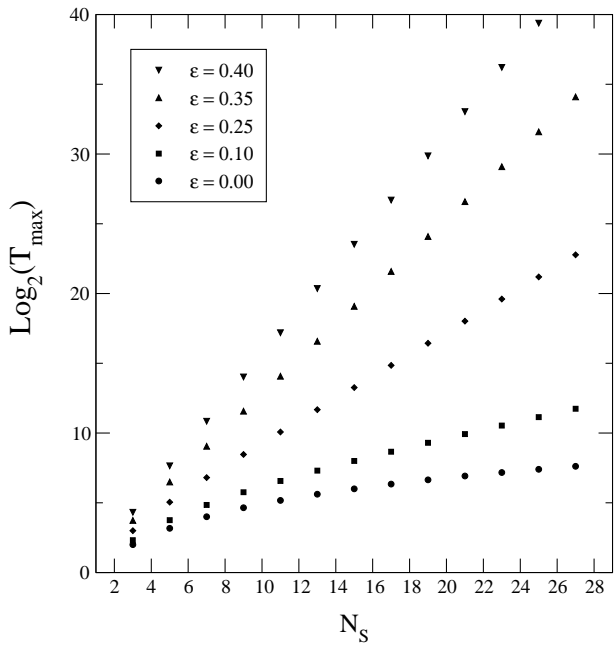}
\caption{\label{tns_aas}
Numerical evaluation of $T_{\mbox{max}}$ as a function of $N_s$ 
for chains with the AA ends and $L$ from 1 to 13.
}
\end{figure}

It is easily seen that the largest value of the MFPT, 
$T_{\mbox{max}}$, is obtained for the RA ends when at all the sites,
the right jump transition is $1/2 - \epsilon$, and the left jump
transition is $1/2 + \epsilon$.
In the asymptotic limit of $N_s \rightarrow \infty$, we obtain
$T_{\mbox{max}} \approx \gamma^{N_s}$~\cite{MKG96}. 
For the AA extremes, $T_{\mbox{max}}$ cannot be easily evaluated. 
Thus, instead of an analytical expression, in Fig.~\ref{tns_aas}
we show a numerical calculation of $T_{\mbox{max}}$ for several
values of $\epsilon$. We see that asymptotically
$\log (T_{\mbox{max}}) \approx N_s$, as in the RA case. 

\begin{figure}
\includegraphics[clip,width=0.45\textwidth]{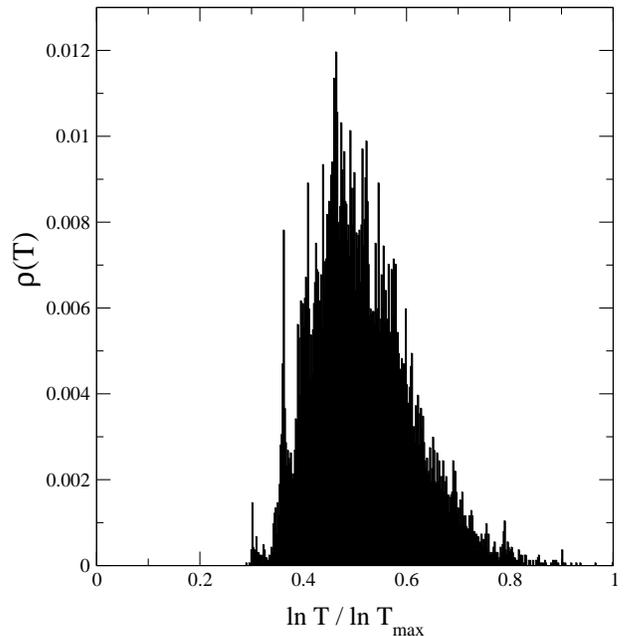}
\caption{\label{haas25}
Histogram of $\rho(T)$ for a chain with $L = 7$, the AA
boundary conditions, and the Sinai disorder with $\epsilon = 0.25$.
We are using the rescaled variable $\ln T / \ln T_{\mbox{max}}$ and
uniform buckets of width equal to $2^9$. $T_{\mbox{max}} = 9824$.
}
\end{figure}

In Fig.~\ref{haas25}, we show the histogram of the distribution
$\rho(T)$ over the dichotomic Sinai disorder. The figure was
constructed computing Eq.~(\ref{MFPT:aa}) for $M=L=7$ ($N_s = 15$),
and using all the possible configurations of disorder.
A related figure, for the histogram of $\rho(T)$ for chains with 
the RA extremes, can be seen in Ref.~\cite{MKG96}.
The distribution $\rho(T)$ has previously been studied~\cite{tail}
for the RA boundary conditions and results broad. 

Our second random biased model for the transition probabilities is
defined by $w^-_j = w^+_j - \epsilon$ and prescribing $w^+_j$ to be
equal to $1/2$ or $3/2$ with equal probabilities. In this case, the
parameter $\epsilon$ measures the strength of the bias, and can take
values between $0$ and $1/2$. In the limit of $\epsilon = 0$, we get a
disordered symmetrical random walk. This dichotomic model corresponds
to a class of weak disorder with global bias.  The quantities
$\beta_k \equiv \left< \left( 1 / w^+_j \right)^k \right>$
result finite for all $k \geq 1$. Therefore, the model does not
present anomalous diffusion~\cite{ABSO81}.
For the MFPT averaged over disorder with the AA boundary conditions,
we obtain up to first order in $\epsilon$~\cite{Pury03,Pury02b},
\begin{equation}
M(1) \simeq
\frac{(L+1)^2}{2 \beta_1^{-1}}
\left[ 1 + \frac{1}{2}
\left( 1 + {\cal F} \right)
\beta_1 \,\epsilon
\right] \,.
\label{<T>:aa_0}
\end{equation}
The asymmetry in the hopping transitions links the strength of the
bias with the fluctuation of the disorder, defined by
${\cal F} \equiv \left(\beta_2 - \beta_1^2 \right) / \beta_1^2$.
In our particular case, we get $\beta_1 = 4/3$, and ${\cal F} = 1/4$.
On the other hand, for the RA extremes, we obtain up to first order
in $\epsilon$~\cite{Pury03},
\begin{equation}
M(1) \simeq
\frac{(L+1)(L+2)}{2 \,\beta_1^{-1}}
\left[ 1 - \frac{L}{3} \beta_1 \,\epsilon \right] \,.
\label{<T>:ra_0}
\end{equation}
Strikingly, for these boundary conditions, the fluctuation
of the disorder is not present in the averaged MFPT.
This fact relies on the difference in the structures of
Eqs.~(\ref{MFPT:aa}) and~(\ref{MFPT:ra_0}).
In the limit of $L \rightarrow \infty$, $M(1$) diverges
as $L^2$ independently of the boundary conditions.

\begin{figure}
\includegraphics[clip,width=0.45\textwidth]{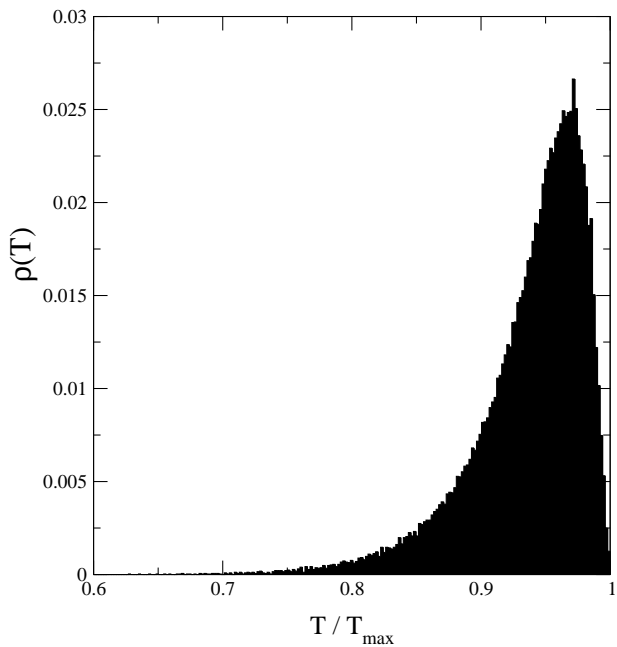}
\caption{\label{haaw25}
Histogram of $\rho(T)$ for a chain with $L = 7$, the AA
boundary conditions, and weak disorder with $\epsilon = 0.25$.
We are using uniform intervals of width equal to $2^9$.
$T_{\mbox{max}} = 31.75$.
}
\end{figure}

\begin{figure}
\includegraphics[clip,width=0.45\textwidth]{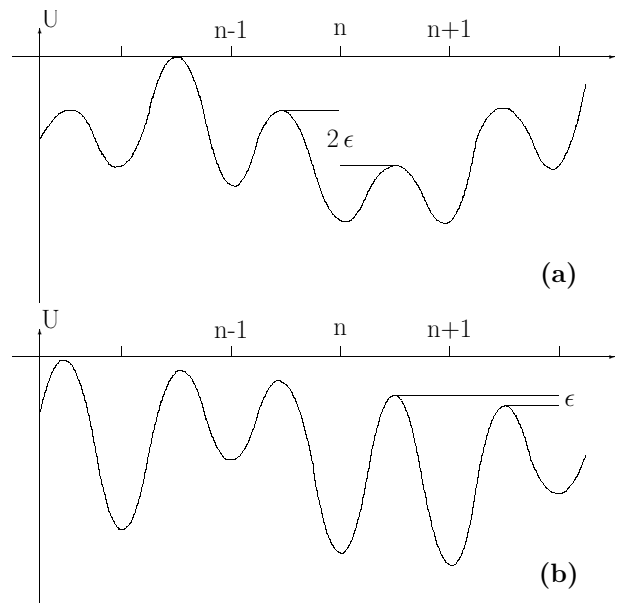}
\caption{\label{lands}
Sketches of the particle potential for both kinds of disorder
and small $\epsilon$:
{\bf (a)} The Sinai model. The difference between two consecutive 
peaks is $2 \epsilon$, with random sign.
{\bf (b)} Weak disorder. The difference between one peak
and the following is always $\epsilon$.
}
\end{figure}

In Fig.~\ref{haaw25}, we show the histogram of the corresponding
distribution $\rho(T)$ over dichotomic weak disorder.
Here, we do not obtain a broad but a more localized distribution.
We must stress the difference in the scales used to construct
the plots in Figs.~\ref{haas25} and~\ref{haaw25}. This fact is
the first indication of the different nature in the distribution
over disorder between both models.
Both models were constructed to get $\left< w^+_j \right> = 1$,
and the main difference among them is in the role of the
parameter $\epsilon$. This parameter controls the disorder in
the Sinai model and regulates the bias in the second model.
Thus, in the Sinai model the bias is local, i.e., the direction
of the bias is randomly drawn in each site, whereas in the
second model, we are considering a global bias field which
points to the right in all sites.
Assuming that the jump probabilities per unit time involve
the Arrhenius factor (see Ref.~\cite{Pury02b}), the landscape
for the particle potential is schematically sketched in
Fig.~\ref{lands} for both models.
%

\section{Multifractal analysis}
\label{sec:multi}

In order to characterize the divergences of the moments of the
MFPT distribution over disorder, we postulate that in the limit 
$N_s \rightarrow \infty$, the $q$th moment obeys the scaling 
relation $M(q) \sim {\cal N}^{\xi(q)}$.
In order to verify this ansatz, we have numerically reckoned
$\log [M(q)]$ vs $N_s$. For chains with the AA ends, we take 
$L$ from 1 to 13 and use Eq.~(\ref{MFPT:aa}) for evaluating the
set $\{T_{(i)}, \,i= 1,\dots, {\cal N} \}$, for a given $L$.
For chains with the RA boundary conditions, we take $L$ from 1 to 20
and use Eq.~(\ref{MFPT:ra_0}) for calculating the values $T_{(i)}$,
for a fixed size of the chain.
For all values of $\epsilon$ and $q$, and for both models of
disorder, with the AA or RA extremes, we obtain straight lines.

\begin{figure}
\includegraphics[clip,width=0.45\textwidth]{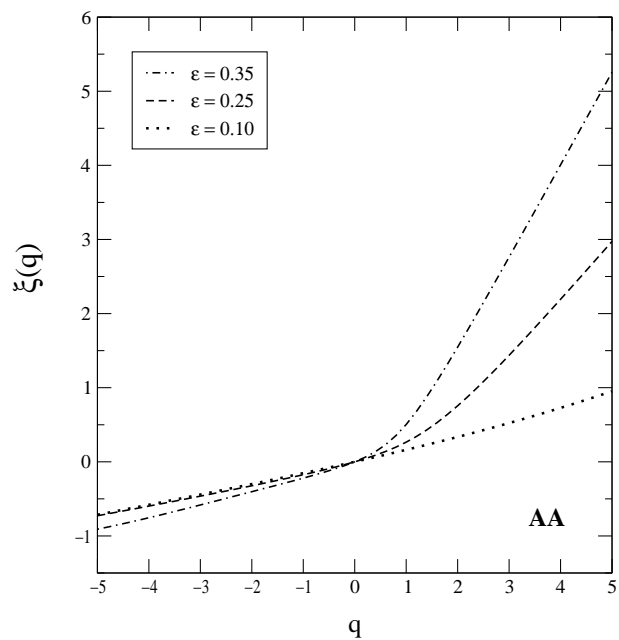}
\caption{\label{xi_aas}
Plot of the scaling exponents $\xi(q)$ for chains with
the AA boundary conditions and the Sinai disorder.
}
\end{figure}

\begin{figure}
\includegraphics[clip,width=0.45\textwidth]{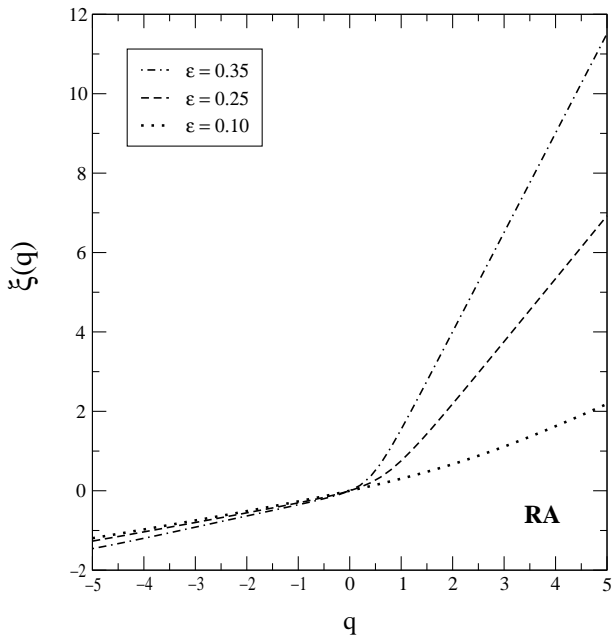}
\caption{\label{xi_ras}
Plot of the scaling exponents $\xi(q)$ for chains with 
the RA boundary conditions and Sinai disorder.
}
\end{figure}

The slope of the linear fitting for the Sinai model with the AA (RA)
ends is plotted in Fig.~\ref{xi_aas}~(Fig.~\ref{xi_ras}) as functions
of $q$, for three values of $\epsilon$.
For all $\epsilon$, we obtain that $\xi(q)$ is a monotonically 
increasing function with $\xi(0) = 0$, which is due to the 
normalization condition of the distribution. We observe that for 
large  values of $\epsilon$, $\xi(q)$ is a nonlinear function;
whereas in the limit of weak disorder, $\xi(q)$ becomes a linear
function.  
For the model of weak disorder, we obtain $\xi(q) = \theta \,q$, 
where $\theta > 0$ is a decreasing function of $\epsilon$.
Therefore, for weak disorder, we get a single gap exponent for
describing the moments of the MFPT distribution.
The analysis with both kinds of boundary conditions leads us to
qualitatively similar behaviors for the exponents $\xi(q)$, 
for both models of disorder described in Sec.~\ref{sec:models}.

Now, let us define the partition function as follows~\cite{HJK+86}:
\begin{equation}
Z(q) = \frac{1}
{\left( \displaystyle\sum_{i=1}^{\cal N} \; T_{(i)} \right)^q} \;
\sum_{i=1}^{\cal N} \; T_{(i)}^q \,.
\label{Z(q)}
\end{equation}
Again, we postulate that in the limit $N_s \rightarrow \infty$, 
the partition function obeys a scaling relation. Thus, we write
$Z(q) \sim {\cal N}^{-\tau(q)}$.
It is well known that for nonmultifractal distributions, the exponents
$\tau(q)$ are linear functions on $q$, namely, $\tau(q) = q-1$.
The multifractality appears with a nontrivial dependence of the
scaling exponents on $q$.  
For all values of $\epsilon$ and $q$, and for both models of disorder,
we obtain straight lines when we plot $\log [Z(q)]$ vs $N_s$.
Therefore, the scaling ansatz is verified. 
Here, we have also used $L$ from 1 to 13 for the AA ends,
and $L$ from 1 to 20 for the RA extremes.

\begin{figure}
\includegraphics[clip,width=0.45\textwidth]{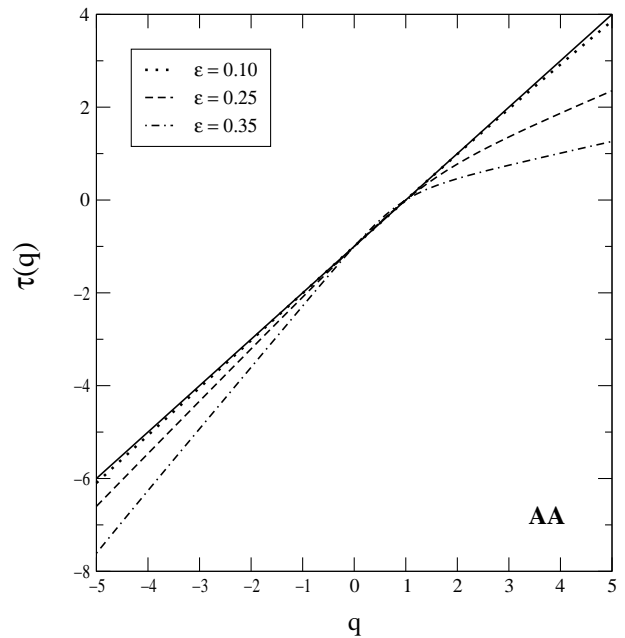}
\caption{\label{tau_aas}
Plot of the scaling exponents $\tau(q)$ for chains with 
the AA boundary conditions and the Sinai disorder, using $L$ 
from 5 to 13.
The solid line corresponds to the linear relation $q-1$.
}
\end{figure}

\begin{figure}
\includegraphics[clip,width=0.45\textwidth]{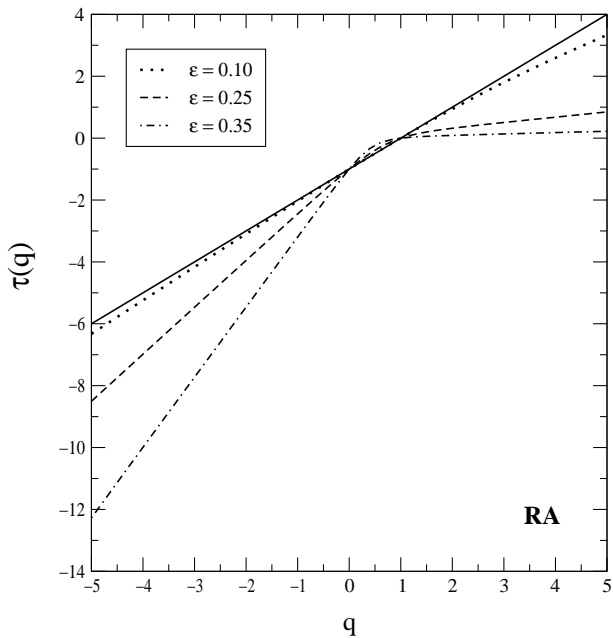}
\caption{\label{tau_ras}
Plot of the scaling exponents $\tau(q)$ for chains with 
the RA boundary conditions and the Sinai disorder, using $L$ 
from 6 to 14.
The solid line corresponds to the linear relation $q-1$.
}
\end{figure}

In Fig.~\ref{tau_aas}~(Fig.~\ref{tau_ras}), we plot $\tau(q)$ 
for chains with the dichotomous Sinai model, the AA (RA) ends, 
and for three representative values of $\epsilon$. 
The functions $\tau(q)$ satisfy $\tau(0) = -1$. 
Additionally, we get $\tau(1) = 0$, 
which is due to the normalization condition of $Z(q)$. 
In the limit of $\epsilon \rightarrow 0$,  $\tau(q) \approx q-1$, 
and for strong disorder, $\tau(q)$ exhibits a nonlinear dependence. 
Thus, the multifractality for the Sinai model is unambiguously
established, independently of the boundary conditions.
For the model of weak disorder, we obtain $\tau(q) = q - 1$ for 
all values of $\epsilon$, within the accuracy of the numerical
evaluations. 
Therefore, we have not found multifractality in our model of weak
disorder, neither with the AA nor with the RA boundary conditions.
This result reinforces the idea that the multifractal phenomenon is
only related to anomalous diffusion. Both models are based on the
same dichotomic rule for assigning values to the transition
probabilities. Moreover, in both cases we found
$\left< w^+_j \right> = 1$. 
In spite of these similar characteristics, the selected models have
quite different behaviors. The random field resulting from the Sinai
condition is at the basis of the anomalous diffusion and the origin
of the multifractal behavior of the MFPT distribution over disorder.

The simple plots obtained in the graphs for the functions $\tau(q)$
suggest us that the multifractality is due to the presence of a
binomial multiplicative process. For this process, the mass exponents
are~\cite{Fed88}
\begin{equation}
\tau(q) = - \frac{\ln \left( p_1^q + p_2^q \right)}{\ln 2} \,,
\label{binomial}
\end{equation}
where $p_1 + p_2 = 1$. This expression immediately satisfies the
conditions $\tau(0) = -1$ and $\tau(1) = 0$, and in the limit
$p_1 = p_2 = 1/2$, results in $\tau(q) = q - 1$. For $p_1 < 1/2$,
we get the limit expressions
\begin{eqnarray}
\tau(q) \rightarrow - q \,\frac{\ln p_1}{\ln 2} &&
\mbox{for } q \rightarrow -\infty  \,, \\
\tau(q) \rightarrow - q \,\frac{\ln p_2}{\ln 2} &&
\mbox{for } q \rightarrow  \infty  \,.
\end{eqnarray}
In our case, the parameters $p_1$ and $p_2$ are functions of the
strength of disorder ($\epsilon$). In Table~\ref{fitted_values},
the values of the fitting of $\tau(q)$, using the two parameters
$p_1$ and $p_2$, are displayed for both kinds of boundary conditions.
\begin{table}
\caption{\label{fitted_values}
Fitted values for the parameters $p_1$ and $p_2$, using
Eq.~(\ref{binomial}), for the plots $\tau(q)$ in
Figs.~\ref{tau_aas} and~\ref{tau_ras}.
}
\begin{ruledtabular}
\begin{tabular}{ccccc}
& \multicolumn{2}{c}{AA} & \multicolumn{2}{c}{RA} \\
$\epsilon$ & $p_1$ & $p_2$ & $p_1$ & $p_2$ \\
\hline
0.00 & 0.500 & 0.500 & 0.500 & 0.500 \\
0.10 & 0.461 & 0.543 & 0.425 & 0.602 \\
0.25 & 0.395 & 0.700 & 0.295 & 0.878 \\
0.35 & 0.336 & 0.825 & 0.173 & 0.964
\end{tabular}
\end{ruledtabular}
\end{table}
We only obtain a good quality of fitting for small values of 
$\epsilon$. For strong disorder, the condition $p_1 + p_2 = 1$ 
is relaxed and we can only fit both asymptotic regimes of the 
curves for $q \rightarrow \pm \infty$.
This fact is a strong indication that the nature of the multifractal
phenomenon is more complex than a multiplicative rule.

\begin{figure}
\includegraphics[clip,width=0.45\textwidth]{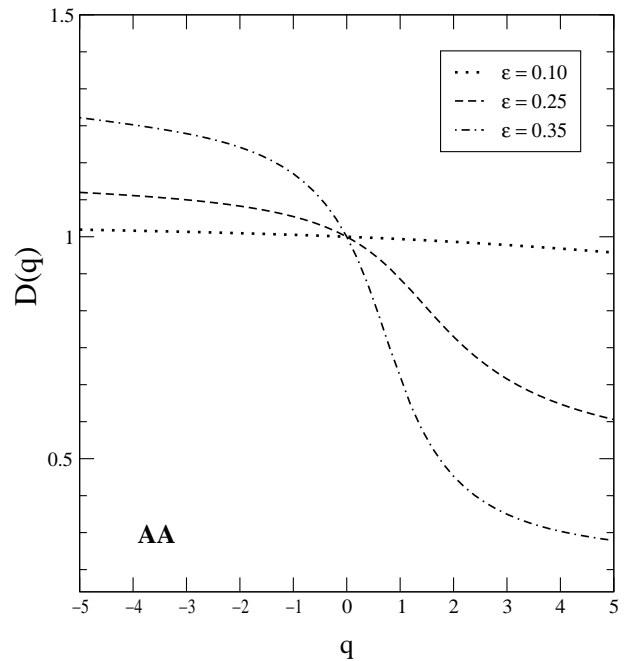}
\caption{\label{d_aas}
Generalized dimensions $D(q)$ for chains with 
the AA boundary conditions and the Sinai disorder.
}
\end{figure}

\begin{figure}
\includegraphics[clip,width=0.45\textwidth]{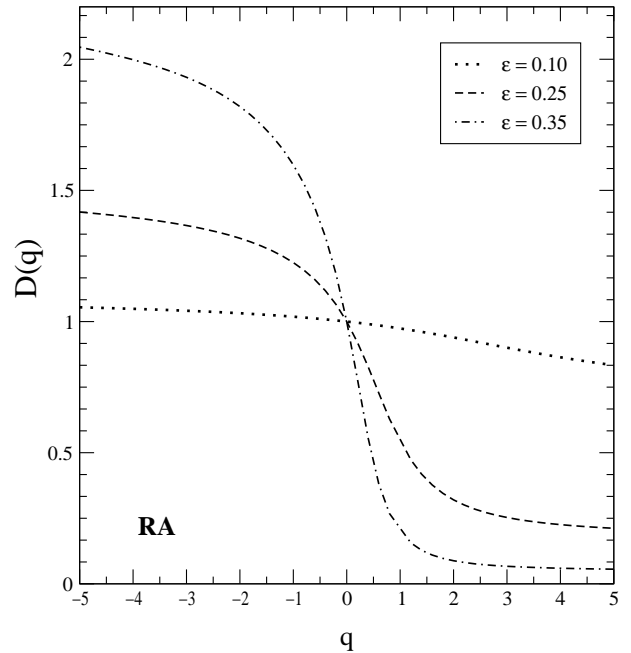}
\caption{\label{d_ras}
Generalized dimensions $D(q)$ for chains with 
the RA boundary conditions and the Sinai disorder.
}
\end{figure}

A usual characterization of multifractality is the generalized Renyi
dimension spectrum~\cite{Fed88} defined by
\footnote{Note that our definition of $\tau(q)$ is equal to minus the
mass exponents used in Ref.~\cite{Fed88}. Usually, the mass exponents
are defined in the limit of a scaling parameter going to zero. 
In our case, the scaling relations hold for the parameter 
${\cal N} \rightarrow \infty$. Therefore, we use the denominator 
$(q-1)$ in the definition of $D(q)$.}
\begin{equation}
D(q) \equiv \frac{\tau(q)}{q-1} \,.
\label{D(q)}
\end{equation}
In Fig.~\ref{d_aas}~(Fig.~\ref{d_ras}) we depict the generalized
dimension spectra for the same conditions and values of $\epsilon$ 
as given by Fig.~\ref{tau_aas}~(Fig.~\ref{tau_ras}). 
We find that $D(q)$ is a monotonically decreasing function of $q$, 
and satisfies $D(0) = 1$, which is the dimension of the support 
of the distribution. 

\begin{figure}
\includegraphics[clip,width=0.45\textwidth]{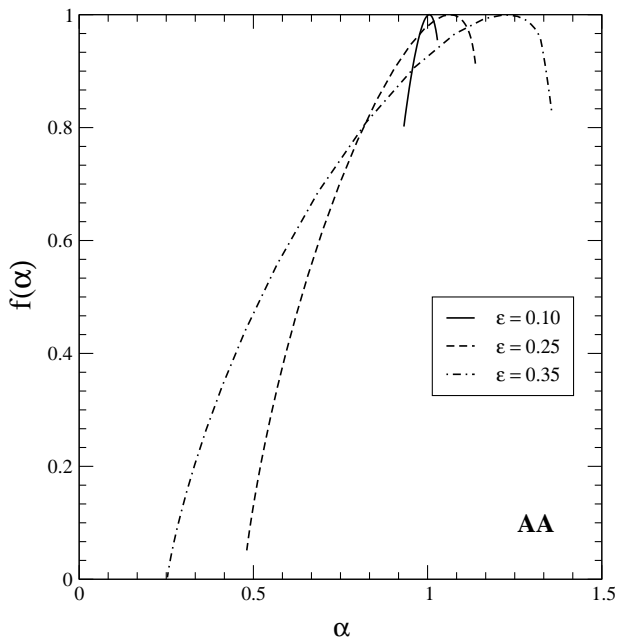}
\caption{\label{f_aas}
Multifractal spectra $f(\alpha)$ for chains with 
the AA boundary conditions and the Sinai disorder.
}
\end{figure}

\begin{figure}
\includegraphics[clip,width=0.45\textwidth]{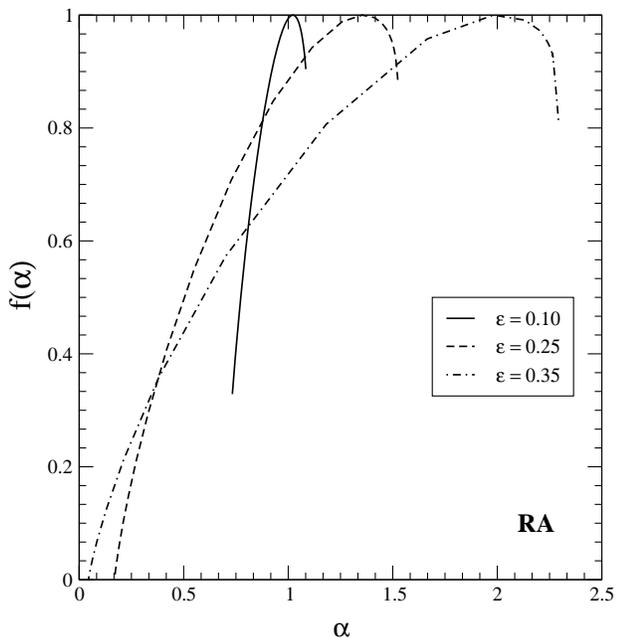}
\caption{\label{f_ras}
Multifractal spectra $f(\alpha)$ for chains with 
the RA boundary conditions and the Sinai disorder.
}
\end{figure}

Taking the Legendre transform of $\tau(q)$, we can obtain the
multifractal spectrum $f(\alpha) = - \tau(q) + q \,\alpha$,
where the Lipschitz-H{\"o}lder exponent $\alpha$ is the derivative 
of $\tau(q)$ with respect to $q$. 
Figure~\ref{f_aas}~(Fig.~\ref{f_ras}) shows the $f(\alpha)$ spectra
for the same conditions and values of $\epsilon$ as given by
Fig.~\ref{tau_aas}~(Fig.~\ref{tau_ras}).
For strong disorder, the curve $f(\alpha)$ becomes broad, whereas for
$\epsilon \rightarrow 0$ the spectrum collapses to the point $(1,1)$.
As is known, the maximum value of $f$ is the fractal dimension of the
support of the measure.

In this work, we have chosen the scaling parameter ${\cal N}$ to
characterize the system size. In Ref.~\cite{MKG96}, a similar
multifractal analysis for the Sinai disorder and the RA boundary
conditions was based on the scaling parameter $T_{\mbox{max}}$.
We have seen in Sec.~\ref{sec:models} that 
$T_{\mbox{max}} \propto {\cal N}$. 
However, this choice of the scaling parameter leads to spurious
results, such as the variation of $D(0)$ with $\epsilon$. 
This phenomenon is an artifact of the mathematical selection of the
scaling parameter, and not a characteristic property of the system
under analysis.  
On the other hand, for one-dimensional systems, we expect that the
dimension of the support of the multifractal measure is equal to 1.

Finally, we derive the relation between the exponents $\xi(q)$ 
and $\tau(q)$. From the definitions given by Eqs.~(\ref{M(q)}) 
and~(\ref{Z(q)}), we inmediatly obtain
\begin{equation}
Z(q) = \frac{{\cal N}^{1-q}}{M(1)^q} \, M(q) \,.
\label{Z(M)}
\end{equation}
Using the scaling ansatz in the last equation and taking logarithms,
we get
\begin{equation}
\tau(q) = q - 1 - \xi(q) + q \,\xi(1) \,.
\label{tau(xi)}
\end{equation}
Particularly, the last relation satisfies inmediately that
$\tau(0) = -1$ and $\tau(1) = 0$.
From the result quoted in Sec.~\ref{sec:models}, for chains with 
the RA ends, we found $\xi(1) = \log_2 \beta(\epsilon)$.

\section{Concluding remarks}
\label{sec:fin}

In this work, we have considered the distribution of the MFPT over 
two classes of disorder, the Sinai and another dichotomic model with 
global bias and nonanomalous diffusion.
Our results confirm us that the multifractality is related
only to anomalous diffusion.

The multifractal behavior in the MFPT distribution over the Sinai
disorder is not a consequence of the multiplicative structure 
of Eq.~(\ref{MFPT:aa}) or~(\ref{MFPT:ra_0}) (the expressions 
for the MFPT for a fixed  realization of disorder), as was 
stated in a previous report~\cite{MKG96}.
The multifractality is an inherent attribute of the strong disorder 
in the Sinai model, and is well established for both kinds of 
boundary conditions.
Moreover, the multifractal signature found in the spectra obtained
suggests us that the origin of the phenomenon is more complex than 
a binomial multiplicative process.


\begin{acknowledgments}
This work was partially supported by grant from
``Se\-cre\-ta\-r\'\i a de Cien\-cia y Tec\-no\-lo\-g\'\i a de la
Uni\-ver\-si\-dad Na\-cio\-nal de C\'or\-doba''
(Code: 05/B160, Res.\ SeCyT 194/00).
P.A.P.\ thanks Centro At\'omico Bariloche for its hospitality
during his stay.
P.A.P.\ is also grateful to Peter Pfeifer for a useful discussion.
\end{acknowledgments}

%


\begin{thebibliography}{26}
\expandafter\ifx\csname natexlab\endcsname\relax\def\natexlab#1{#1}\fi
\expandafter\ifx\csname bibnamefont\endcsname\relax
  \def\bibnamefont#1{#1}\fi
\expandafter\ifx\csname bibfnamefont\endcsname\relax
  \def\bibfnamefont#1{#1}\fi
\expandafter\ifx\csname citenamefont\endcsname\relax
  \def\citenamefont#1{#1}\fi
\expandafter\ifx\csname url\endcsname\relax
  \def\url#1{\texttt{#1}}\fi
\expandafter\ifx\csname urlprefix\endcsname\relax\def\urlprefix{URL }\fi
\providecommand{\bibinfo}[2]{#2}
\providecommand{\eprint}[2][]{\url{#2}}

\bibitem[{\citenamefont{Alexander et~al.}(1981)
\citenamefont{Alexander,Bernasconi, Schneider, and Orbach}}]{ABSO81}
  \bibinfo{author}{\bibfnamefont{S.}~\bibnamefont{Alexander}},
  \bibinfo{author}{\bibfnamefont{J.}~\bibnamefont{Bernasconi}},
  \bibinfo{author}{\bibfnamefont{W.~R.} \bibnamefont{Schneider}},
  \bibnamefont{and}
  \bibinfo{author}{\bibfnamefont{R.}~\bibnamefont{Orbach}},
  \bibinfo{journal}{Rev.\ Mod.\ Phys.} \textbf{\bibinfo{volume}{53}},
  \bibinfo{pages}{175} (\bibinfo{year}{1981}).

\bibitem{Reviews}
  \bibinfo{author}{\bibfnamefont{J.~W.} \bibnamefont{Haus}}
  \bibnamefont{and}
  \bibinfo{author}{\bibfnamefont{K.~W.} \bibnamefont{Kehr}},
  \bibinfo{journal}{Phys.\ Rep.} \textbf{\bibinfo{volume}{150}},
  \bibinfo{pages}{263} (\bibinfo{year}{1987});
%
  \bibinfo{author}{\bibfnamefont{S.}~\bibnamefont{Havlin}}
  \bibnamefont{and}
  \bibinfo{author}{\bibfnamefont{D.}~\bibnamefont{Ben-Avraham}},
  \bibinfo{journal}{Adv.\ Phys.} \textbf{\bibinfo{volume}{36}},
  \bibinfo{pages}{695} (\bibinfo{year}{1987});
%
  \bibinfo{author}{\bibfnamefont{J.~P.} \bibnamefont{Bouchaud}}
  \bibnamefont{and}
  \bibinfo{author}{\bibfnamefont{A.}~\bibnamefont{Georges}},
  \bibinfo{journal}{Phys.\ Rep.} \textbf{\bibinfo{volume}{195}},
  \bibinfo{pages}{127} (\bibinfo{year}{1990}).

\bibitem[{\citenamefont{Havlin and Bunde}(1991)}]{HB91}
  \bibinfo{author}{\bibfnamefont{S.}~\bibnamefont{Havlin}}
  \bibnamefont{and}
  \bibinfo{author}{\bibfnamefont{A.}~\bibnamefont{Bunde}}, in
  \emph{\bibinfo{booktitle}{Fractals and Disordered Systems}},
  edited by
  \bibinfo{editor}{\bibfnamefont{A.}~\bibnamefont{Bunde}}
  \bibnamefont{and}
  \bibinfo{editor}{\bibfnamefont{S.}~\bibnamefont{Havlin}}
  (\bibinfo{publisher}{Springer-Verlag},
  \bibinfo{address}{Berlin},
  \bibinfo{year}{1991}), \bibinfo{note}{Chap.\ 3}.

\bibitem[{\citenamefont{Bunde et~al.}(1987)
  \citenamefont{Bunde, Harder, Havlin,and Roman}}]{BHHR87}
  \bibinfo{author}{\bibfnamefont{A.}~\bibnamefont{Bunde}},
  \bibinfo{author}{\bibfnamefont{H.}~\bibnamefont{Harder}},
  \bibinfo{author}{\bibfnamefont{S.}~\bibnamefont{Havlin}},
  \bibnamefont{and}
  \bibinfo{author}{\bibfnamefont{H.~E.} \bibnamefont{Roman}},
  \bibinfo{journal}{J.\ Phys.\ A} \textbf{\bibinfo{volume}{20}},
  \bibinfo{pages}{L865} (\bibinfo{year}{1987}).

\bibitem[{\citenamefont{Roman et~al.}(1988)
  \citenamefont{Roman, Bunde, and Havlin}}]{RBH88}
  \bibinfo{author}{\bibfnamefont{H.~E.} \bibnamefont{Roman}},
  \bibinfo{author}{\bibfnamefont{A.}~\bibnamefont{Bunde}},
  \bibnamefont{and}
  \bibinfo{author}{\bibfnamefont{S.}~\bibnamefont{Havlin}},
  \bibinfo{journal}{Phys.\ Rev.\ A} \textbf{\bibinfo{volume}{38}},
  \bibinfo{pages}{2185} (\bibinfo{year}{1988}).

\bibitem[{\citenamefont{Bunde et~al.}(1990)
  \citenamefont{Bunde, Havlin, and Roman}}]{BHR90}
  \bibinfo{author}{\bibfnamefont{A.}~\bibnamefont{Bunde}},
  \bibinfo{author}{\bibfnamefont{S.}~\bibnamefont{Havlin}},
  \bibnamefont{and}
  \bibinfo{author}{\bibfnamefont{H.~E.} \bibnamefont{Roman}},
  \bibinfo{journal}{Phys.\ Rev.\ A} \textbf{\bibinfo{volume}{42}},
  \bibinfo{pages}{6274} (\bibinfo{year}{1990}).

\bibitem[{\citenamefont{Roman}(1992)}]{Rom92}
\bibinfo{author}{\bibfnamefont{H.~E.} \bibnamefont{Roman}},
  \bibinfo{journal}{Physica A} \textbf{\bibinfo{volume}{191}},
  \bibinfo{pages}{379} (\bibinfo{year}{1992}).

\bibitem[{\citenamefont{Eisenberg et~al.}(1993)
  \citenamefont{Eisenberg, Bunde, Havlin, and Roman}}]{EBHR93}
  \bibinfo{author}{\bibfnamefont{E.}~\bibnamefont{Eisenberg}},
  \bibinfo{author}{\bibfnamefont{A.}~\bibnamefont{Bunde}},
  \bibinfo{author}{\bibfnamefont{S.}~\bibnamefont{Havlin}}, 
  \bibnamefont{and}
  \bibinfo{author}{\bibfnamefont{H.~E.} \bibnamefont{Roman}},
  \bibinfo{journal}{Phys.\ Rev.\ E} \textbf{\bibinfo{volume}{47}},
  \bibinfo{pages}{2333} (\bibinfo{year}{1993}).

\bibitem[{\citenamefont{Eisenberg et~al.}(1994)
  \citenamefont{Eisenberg, Havlin, and Weiss}}]{EHW94}
  \bibinfo{author}{\bibfnamefont{E.}~\bibnamefont{Eisenberg}},
  \bibinfo{author}{\bibfnamefont{S.}~\bibnamefont{Havlin}}, 
  \bibnamefont{and}
  \bibinfo{author}{\bibfnamefont{G.~H.} \bibnamefont{Weiss}},
  \bibinfo{journal}{Phys.\ Rev.\ Lett.} \textbf{\bibinfo{volume}{72}},
  \bibinfo{pages}{2827} (\bibinfo{year}{1994}).

\bibitem[{\citenamefont{Ma\v{s}i\'c and Djordjevi\'c}(1990)}]{MD90}
  \bibinfo{author}{\bibfnamefont{N.}~\bibnamefont{Ma\v{s}i\'c}} 
  \bibnamefont{and}
  \bibinfo{author}{\bibfnamefont{Z.~V.} \bibnamefont{Djordjevi\'c}},
  \bibinfo{journal}{Physica A} \textbf{\bibinfo{volume}{167}},
  \bibinfo{pages}{560} (\bibinfo{year}{1990}).

\bibitem[{\citenamefont{den Broeck}(1991)}]{van91}
  \bibinfo{author}{\bibfnamefont{C.~Van} \bibnamefont{den Broeck}},
  \bibinfo{journal}{J.\ Stat.\ Phys.} \textbf{\bibinfo{volume}{65}},
  \bibinfo{pages}{971} (\bibinfo{year}{1991}).

\bibitem[{\citenamefont{Murthy et~al.}(1994)
  \citenamefont{Murthy, Rajasekar, and Kehr}}]{MRK94}
  \bibinfo{author}{\bibfnamefont{K.~P.~N.} \bibnamefont{Murthy}},
  \bibinfo{author}{\bibfnamefont{S.}~\bibnamefont{Rajasekar}},
  \bibnamefont{and} 
  \bibinfo{author}{\bibfnamefont{K.~W.} \bibnamefont{Kehr}},
  \bibinfo{journal}{J.\ Phys.\ A} \textbf{\bibinfo{volume}{27}},
  \bibinfo{pages}{L107} (\bibinfo{year}{1994}).

\bibitem[{\citenamefont{Wichmann et~al.}(1995)
  \citenamefont{Wichmann, Giacometti, and Murthy}}]{WGM95}
  \bibinfo{author}{\bibfnamefont{T.}~\bibnamefont{Wichmann}},
  \bibinfo{author}{\bibfnamefont{A.}~\bibnamefont{Giacometti}},
  \bibnamefont{and} 
  \bibinfo{author}{\bibfnamefont{K.~P.~N.} \bibnamefont{Murthy}}, 
  \bibinfo{journal}{Phys.\ Rev.\ E} \textbf{\bibinfo{volume}{52}},
  \bibinfo{pages}{481} (\bibinfo{year}{1995}).

\bibitem[{\citenamefont{Murthy et~al.}(1996{\natexlab{a}})
  \citenamefont{Murthy, Kehr, and Giacometti}}]{MKG96}
  \bibinfo{author}{\bibfnamefont{K.~P.~N.} \bibnamefont{Murthy}},
  \bibinfo{author}{\bibfnamefont{K.~W.} \bibnamefont{Kehr}}, 
  \bibnamefont{and}
  \bibinfo{author}{\bibfnamefont{A.}~\bibnamefont{Giacometti}},
  \bibinfo{journal}{Phys.\ Rev.\ E} \textbf{\bibinfo{volume}{53}},
  \bibinfo{pages}{444} (\bibinfo{year}{1996}{\natexlab{a}}).

\bibitem[{\citenamefont{Murthy et~al.}(1996{\natexlab{b}})
  \citenamefont{Murthy, Giacometti, and Kehr}}]{MGK96}
  \bibinfo{author}{\bibfnamefont{K.~P.~N.} \bibnamefont{Murthy}},
  \bibinfo{author}{\bibfnamefont{A.}~\bibnamefont{Giacometti}},
  \bibnamefont{and} 
  \bibinfo{author}{\bibfnamefont{K.~W.} \bibnamefont{Kehr}},
  \bibinfo{journal}{Physica A} \textbf{\bibinfo{volume}{224}},
  \bibinfo{pages}{232} (\bibinfo{year}{1996}{\natexlab{b}}).

\bibitem{Kim}
  \bibinfo{author}{\bibfnamefont{K.}~\bibnamefont{Kim}},
  \bibinfo{author}{\bibfnamefont{J.~S.} \bibnamefont{Choi}}, 
  \bibnamefont{and}
  \bibinfo{author}{\bibfnamefont{Y.~S.} \bibnamefont{Kong}},
  \bibinfo{journal}{J.\ Phys.\ Soc.\ Jpn.} \textbf{\bibinfo{volume}{67}},
  \bibinfo{pages}{1583} (\bibinfo{year}{1998});
%
  \bibinfo{author}{\bibfnamefont{K.}~\bibnamefont{Kim}},
  \bibinfo{author}{\bibfnamefont{G.~H.} \bibnamefont{Kim}}, 
  \bibnamefont{and}
  \bibinfo{author}{\bibfnamefont{Y.~S.} \bibnamefont{Kong}},
  \bibinfo{journal}{Fractals} \textbf{\bibinfo{volume}{8}},
  \bibinfo{pages}{181} (\bibinfo{year}{2000}).

\bibitem[{\citenamefont{Halsey et~al.}(1986)
  \citenamefont{Halsey, Jensen, Procaccia, and Shraiman}}]{HJK+86}
  \bibinfo{author}{\bibfnamefont{T.~C.} \bibnamefont{Halsey}},
  \bibinfo{author}{\bibfnamefont{M.~H.} \bibnamefont{Jensen}},
  \bibinfo{author}{\bibfnamefont{L.~P.} \bibnamefont{Kadanoff}},
  \bibinfo{author}{\bibfnamefont{I.} \bibnamefont{Procaccia}},
  \bibnamefont{and} 
  \bibinfo{author}{\bibfnamefont{B.~I.} \bibnamefont{Shraiman}}, 
  \bibinfo{journal}{Phys.\ Rev.\ A} \textbf{\bibinfo{volume}{33}},
  \bibinfo{pages}{1141} (\bibinfo{year}{1986}).

\bibitem[{\citenamefont{Feder}(1988)}]{Fed88}
\bibinfo{author}{\bibfnamefont{J.}~\bibnamefont{Feder}},
  \emph{\bibinfo{title}{Fractals}} (\bibinfo{publisher}{Plenum Press},
  \bibinfo{address}{New York}, \bibinfo{year}{1988}).

\bibitem{Gallos}
  \bibinfo{author}{\bibfnamefont{L.~K.} \bibnamefont{Gallos}},
  \bibinfo{author}{\bibfnamefont{P.}~\bibnamefont{Argyrakis}},
  \bibnamefont{and} 
  \bibinfo{author}{\bibfnamefont{K.~W.} \bibnamefont{Kehr}},
  \bibinfo{journal}{Phys.\ Rev.\ E} \textbf{\bibinfo{volume}{52}},
  \bibinfo{pages}{1520} (\bibinfo{year}{1995});
%
  \bibinfo{author}{\bibfnamefont{K.~P.~N.} \bibnamefont{Murthy}},
  \bibinfo{author}{\bibfnamefont{L.~K.} \bibnamefont{Gallos}},
  \bibinfo{author}{\bibfnamefont{P.}~\bibnamefont{Argyrakis}},
  \bibnamefont{and} 
  \bibinfo{author}{\bibfnamefont{K.~W.} \bibnamefont{Kehr}},
  \bibinfo{journal}{{\em ibid.}} \textbf{\bibinfo{volume}{54}},
  \bibinfo{pages}{6922} (\bibinfo{year}{1996}).

\bibitem[{\citenamefont{Sinai}(1982)}]{sinai}
  \bibinfo{author}{\bibfnamefont{Y.~G.} \bibnamefont{Sinai}},
  \bibinfo{journal}{Theor.\ Probab.\ Appl.} \textbf{\bibinfo{volume}{27}},
  \bibinfo{pages}{256} (\bibinfo{year}{1982}).

\bibitem[{\citenamefont{Harisha and Murthy}(2000)}]{HM00}
  \bibinfo{author}{\bibfnamefont{R.}~\bibnamefont{Harisha}} 
  \bibnamefont{and}
  \bibinfo{author}{\bibfnamefont{K.}~\bibnamefont{Murthy}},
  \bibinfo{journal}{Physica A} \textbf{\bibinfo{volume}{287}},
  \bibinfo{pages}{161} (\bibinfo{year}{2000}).

\bibitem[{\citenamefont{Pury and C{\'a}ceres}(2003)}]{Pury03}
  \bibinfo{author}{\bibfnamefont{P.~A.} \bibnamefont{Pury}} 
  \bibnamefont{and}
  \bibinfo{author}{\bibfnamefont{M.~O.} \bibnamefont{C{\'a}ceres}},
  \bibinfo{journal}{J.\ Phys.\ A} \textbf{\bibinfo{volume}{36}},
  \bibinfo{pages}{2695} (\bibinfo{year}{2003}).

\bibitem[{\citenamefont{Murthy and Kehr}(1989)}]{MK89}
  \bibinfo{author}{\bibfnamefont{K.~P.~N.} \bibnamefont{Murthy}}
  \bibnamefont{and} 
  \bibinfo{author}{\bibfnamefont{K.~W.} \bibnamefont{Kehr}},
  \bibinfo{journal}{Phys.\ Rev.\ A} \textbf{\bibinfo{volume}{40}},
  \bibinfo{pages}{2082} (\bibinfo{year}{1989});
  \bibinfo{note}{{\bf 41}, 1160(E) (1990)}.

\bibitem[{\citenamefont{Noskowicz and Goldhirsch}(1988)}]{NG88}
  \bibinfo{author}{\bibfnamefont{S.~H.} \bibnamefont{Noskowicz}}
  \bibnamefont{and}
  \bibinfo{author}{\bibfnamefont{I.}~\bibnamefont{Goldhirsch}},
  \bibinfo{journal}{Phys.\ Rev.\ Lett.} \textbf{\bibinfo{volume}{61}},
  \bibinfo{pages}{500} (\bibinfo{year}{1988}).

\bibitem{tail}
  \bibinfo{author}{\bibfnamefont{K.~W.} \bibnamefont{Kehr}} 
  \bibnamefont{and}
  \bibinfo{author}{\bibfnamefont{K.~P.~N.} \bibnamefont{Murthy}},
  \bibinfo{journal}{Phys.\ Rev.\ A} \textbf{\bibinfo{volume}{41}},
  \bibinfo{pages}{5728} (\bibinfo{year}{1990});
%
  \bibinfo{author}{\bibfnamefont{S.~H.} \bibnamefont{Noskowicz}}
  \bibnamefont{and}
  \bibinfo{author}{\bibfnamefont{I.}~\bibnamefont{Goldhirsch}},
  \bibinfo{journal}{{\em ibid.}} \textbf{\bibinfo{volume}{42}},
  \bibinfo{pages}{2047} (\bibinfo{year}{1990}).

\bibitem[{\citenamefont{Pury and C{\'a}ceres}(2002)}]{Pury02b}
  \bibinfo{author}{\bibfnamefont{P.~A.} \bibnamefont{Pury}} 
  \bibnamefont{and}
  \bibinfo{author}{\bibfnamefont{M.~O.} \bibnamefont{C{\'a}ceres}},
  \bibinfo{journal}{Phys.\ Rev.\ E} \textbf{\bibinfo{volume}{66}},
  \bibinfo{pages}{021112} (\bibinfo{year}{2002}).

\end{thebibliography}

\end{document}